\documentclass[reprint,eqsecnum,floats,aps,amsmath,amssymb,footinbib,prd,twocolumn, showpacs]{revtex4-1}
\usepackage{amsmath,amsthm,latexsym,amssymb,amsfonts}

\begin{document}

\title{Uniqueness of the Fock quantization of scalar fields and processes with signature change in cosmology}
\author{Laura Castell\'o Gomar}
\email{laura.castello@iem.cfmac.csic.es}
\author{Guillermo A. Mena Marug\'an}
\email{mena@iem.cfmac.csic.es}
\affiliation{Instituto de Estructura de la Materia, IEM-CSIC, Serrano 121, 28006 Madrid, Spain}

\begin{abstract}
We study scalar fields subject to an equation of the Klein-Gordon type in nonstationary spacetimes, such as those found in cosmology, assuming that all the relevant spatial dependence is contained in the Laplacian. We show that the field description ---with a specific canonical pair--- and the Fock representation for the quantization of the field are fixed indeed in a unique way (except for unitary transformations that do not affect the physical predictions) if we adopt the combined criterion of (a) imposing the invariance of the vacuum under the group of spatial symmetries of the field equations and (b) requiring a unitary implementation of the dynamics in the quantum theory. Besides, we provide a spacetime interpretation of the field equations as those corresponding to a scalar field in a cosmological spacetime that is conformally ultrastatic. In addition, in the privileged Fock quantization, we investigate the generalization of the evolution of physical states from the hyperbolic dynamical regime to an elliptic regime. In order to do this, we contemplate the possibility of processes with signature change in the spacetime where the field propagates and discuss the behavior of the background geometry when the change happens, proving that the spacetime metric degenerates. Finally, we argue that this kind of signature change leads naturally to a phenomenon of particle creation, with exponential production.
\end{abstract}


\maketitle

\section{Introduction}

A challenge for Modern Physics is to incorporate the quantum character of the laws of nature by quantizing the theories that describe the Universe in a classical regime. Ensuring the uniqueness of this procedure is essential to avoid ambiguities that affect the robustness of physical predictions. As opposed to the situation found in Quantum Mechanics (where this is guaranteed by the Stone-von Neumann uniqueness theorem \cite{simon}), linear canonical transformations are not generally implemented as unitary transformations in Quantum Field Theory (QFT) \cite{wald}. Consequently, this introduces an ambiguity in the selection of the Fock representation for the canonical commutation relations, understandable as the freedom in the choice of inequivalent vacua. Different choices of creation and annihilationlike variables are related by linear canonical transformations which cannot be all implemented unitarily under quantization. Moreover, there exists a vast collection of possible canonical pairs of variables to describe the field. In particular, in cosmological scenarios the nonstationarity of the spacetime leads to the possibility of absorbing part of the time dependence of the field via its scaling by a function that depends only on the background. Then, one can rewrite the equation of motion as that for a field that propagates in a static spacetime with a topology identical to the original one, but in return a time dependent mass term emerges \cite{jcap1}. For instance, this is the situation found when one considers massive fields in Friedmann-Robertson-Walker (FRW) universes, as well as in the case of the de Sitter spacetime.

For fields propagating in highly symmetric spacetimes, the invariance under the isometries of the background is enough to select a unique Fock quantization \cite{wald,AshMagFlor}. The picture gets much more complicated when one deals with more realistic scenarios, inasmuch as there is not sufficient symmetry to solve the ambiguity. However, in general cases, the symmetry invariance remains a useful tool in order to select a preferred Fock representation when additional requirements are introduced. Along these lines, in general systems whose linear field equations have no timelike symmetries, it seems reasonable to demand unitarity in the time evolution, as the next weaker possibility instead of invariance. Unitarity is crucial if one is not willing to renounce to a standard probabilistic interpretation of the quantum theory. Recently, the generalization of this criterion has been proven for spatial sections with any compact topology in three or less dimensions \cite{generalproof1}. Remarkably, the two commented types of ambiguities (i.e., those in the choice of canonical pair and in the choice of Fock representation) are removed by the combined requirement of (a) invariance of the vacuum under the spatial symmetries of the field equation, and (b) unitary implementation of the dynamics in the quantum theory. This result has been rigorously demonstrated \cite{generalproof1} for a scalar field that satisfies a Klein-Gordon (KG) equation of the form:
\begin{equation}\label{KG}
\varphi^{\prime\prime}-\Delta\varphi + m^2(t)\varphi =0,
\end{equation}
where the prime stands for the derivative with respect to the time $t$, $\Delta$ is the Laplacian, or more precisely the Laplace-Beltrami (LB) operator, and $m(t)$ is a rather generic real mass function, which must only satisfy some extremely mild conditions (that we will specify later in Sec. II). In this privileged quantization, the canonical field momentum is chosen as the time derivative of the field configuration, properly densitized.

This uniqueness criterion was first proven for the case of linearly polarized Gowdy models \cite{GowMod}. Additionally, a context in which the discussion encounters a natural application is in the quantization of cosmological perturbations; for instance, with the introduction of Mukhanov-Sasaki variables for the scalar perturbations around FRW \cite{MukhaSaki}. The tensor perturbations of an FRW cosmological background, describing its gravitational wave content, are subject as well to a field equation of this type after scaling them (and choosing conformal time) \cite{Bardeen}.

One of the goals of the present article is to extend this criterion to the most general possible field equation, so that its application can be generalized to a broader class of physical scenarios. In order to do this, we consider a generic second-order differential equation of motion of generalized KG type, introducing not only a time dependent scaling of the field, but allowing also a reparametrization of time. Nonetheless, we assume a certain restriction on the spatial dependence, so that the spatial variation appears in the field equation only through the LB operator. We will show with these premises that, given a generic equation of motion of the considered type, it is always possible to find a transformation of the canonical variables leading to a new formulation in which the field satisfies Eq. (\ref{KG}) and admits a unique Fock representation according to the proposed criterion. Indeed, as we have said above, this representation is invariant and allows a unitary implementation of the (transformed) field evolution. Moreover, we will see that the change of canonical variables that relates the original equation with Eq. (\ref{KG}) is univocally determined.

This general result refutes some claims appeared in the literature about QFT in cosmological spacetimes. For instance, in Ref. \cite{edd} it was stated that, for the case of a massless field in de Sitter spacetime, no scaling of the field variable allows a unitary dynamics. In that case, the field equation has a term proportional to the first time derivative and there is a time dependent function that multiplies the Laplacian. Crucial for unitarity is the choice of the conjugate momentum adapted to conformal time. Actually, it was shown in Ref. \cite{dadeSit} that the standard conformal scaling of the field does lead to a formulation with the desired properties if the canonical conjugate momentum includes a linear contribution of the field configuration. In fact, this particular choice of canonical pair is the only one which allows for unitary dynamics, regardless of the value of the mass of the original field in de Sitter spacetime. One of our aims in the present work is to clarify how the uniqueness criterion that we put forward can be extended to situations of this sort, with equations that might be thought to contain damping or friction contributions.

Another aim of this article is to provide a spacetime interpretation of the class of field equations to which we will extend the application of the uniqueness criterion. Among the wide range of imaginable scenarios, we contemplate the possibility of a change of behavior in the field equation from a hyperbolic evolution to an elliptic regime. In the following, we will use the terminology ``signature change'' to describe a variation of this type in the properties of the equations, similar to that experimented when the evolution is analytically continued from a timelike to a spacelike direction, resulting in a change from Lorentzian ($ - + + + $) to Euclidean ($ + + + + $) signature in the metric (e.g. for families of conformally ultrastatic spacetimes).

Signature change phenomena in a general relativity framework have already been explored and discussed from different perspectives in the literature, both with respect to the physical mechanism and its mathematical treatment. The Hartle-Hawking state is one of the most notorious proposals in this sense. In the early $80$'s, Hartle and Hawking introduced the idea of a ``no-boundary'' quantum state, obtained by means of a Euclidean path integral over geometries which do not present other boundary than that corresponding to the observed section of the Universe \cite{noboundary}.  At least in simple models, this state could be interpreted as one that experiences a process in which the signature of the early universe changes from Euclidean to Lorentzian \cite{noboundary}. More generally, nonetheless, the state has to be defined by a path integral over complex integration contours \cite{halliwell} and, in this complexification, the occurrence of a signature change is obscured.

Recently, signature change has reentered the scene within a different quantum approach, appearing in some of the latter works \cite{cambiosignatura,deformedLQC,MS,cai} in loop quantum cosmology (LQC) \cite {LQC,LQCmena} about symmetry reduced models or perturbations around them. We recall that LQC addresses the quantization of cosmological systems following the ideas and methods of loop quantum gravity (LQG) \cite {GCL}, a nonperturbative and background independent formalism designed to quantize Einstein's theory. One of the most remarkable predictions of LQC is the emergence of a quantum bounce mechanism (known as Big Bounce) when the matter density ($\rho$) and the curvature approximate the Planck order. The initial Big Bang singularity is eluded, connecting the bounce with a new evolutive branch in contraction \cite{bounce}. In LQG, the quantization of the spacetime near the Planck scale is manifested through the spectral discretization of the geometric operators, like the area and the volume operators. The effects of this spacetime discretization are introduced by holonomy corrections \cite{holonomy,MS,cai}, as well as by related corrections arising from the regulation of the inverse of the volume \cite{inverse}, both of which are argued to modify the classical constraints \cite{both,ucai}. This procedure suffers from various ambiguities, and the resulting algebra is not closed in general. Actually, there are only certain ways of varying the constraints such that the closure of the algebra is attained \cite{cai}. Nevertheless, not only the classical dynamics changes, but it has been defended that the spacetime structure itself is deformed as a result of these modifications.

For instance, this becomes clear when we consider scalar perturbations in a flat FRW background \cite{MS,ucai}. In that case, the evolution equation for the gauge invariant description $v$ of the scalar perturbations \cite {bardeen}, known as the Mukhanov-Sasaki equation \cite{mukhanov}, has been argued to adopt the form \cite{note1}:
\begin{equation}
\upsilon^{\prime\prime} - \left(1-\frac{2\rho}{\rho_c}\right)\Delta\upsilon -\frac{z^{\prime\prime}}{z}\upsilon =0.
\end{equation}
Here $\rho_c$ is the critical density, at which the bounce occurs in LQC, and $z$ is a function that only depends on the background. We notice that this equation indicates a transition between a hyperbolic behavior for densities $\rho < \rho_c/2$, usually associated with a Lorentzian spacetime regime, and an elliptic behavior for $ \rho > \rho_c/2$, generally thought of as corresponding to a Euclidean regime. Namely, the effective description points out to a transition to a Euclidean spacetime for high energy densities (and large curvatures). The consequences of this phenomenon, if real, are not yet known; nonetheless, we will discuss here how one may deal with it in QFT. The change from a Euclidean spacetime to a Lorentzian one occurs on a spatial hypersurface at a certain time, $t_d$, in a very early epoch of the Universe, distinguishing two spacetime regions with very different properties. Our analysis about the treatment of this signature change will be focused on studying the behavior of geometric quantities across the ``boundary'' hypersurface and on the discussion of the evolution of a vacuum state when passing from one region to another. The last issue that we will explore in this work is the possible production of particles as a result of this process of signature change.

The rest of the paper is organized as follows. In Sec. II, we present a brief summary of the results that have been obtained so far about the uniqueness of the Fock description of a scalar field with a time dependent mass (for any compact topology in no more than three spatial dimensions). A general discussion can be found in Refs. \cite{generalproof1, generalproof2}. It was proven there that the selection of the Fock representation is unique ---up to unitary transformations--- if one demands invariance of the vacuum under the group of spatial symmetries of the field equations and unitary implementability of the dynamics in the quantum theory. In Sec. III, we generalize the class of field equations for which the uniqueness results find applications. This generalization is achieved by considering both time dependent canonical transformations in which the field configuration is scaled and/or time reparametrizations. A spacetime interpretation is then provided in Sec. IV for all those field equations. Moreover, we show that, under certain conditions, there may be a transition process from a Euclidean spacetime region to another of Lorentzian type, and we discuss the applicability of our results in these circumstances. In Sec. V, we study the evolution of a vacuum state in a spacetime with a signature change of these characteristics, assuming that the vacuum is defined by certain conditions at an ``initial time'' in the Euclidean region. Finally, we summarize the main results of this work in Sec. VI.

\section{Uniqueness of the Fock quantization}

We will first describe succinctly the uniqueness results which select a Fock representation (up to unitary transformations) of a KG field with a time dependent mass in a generic ultrastatic background with compact spatial sections. Compactness is required to avoid complications in the infrared region (because then the spectrum of the LB operator is discrete, with a finite number of zero modes). The requirement of compactness is not physically relevant inasmuch as scales beyond the Hubble radius should not be significant for cosmology. Detailed proofs can be found in Ref. \cite{generalproof1, generalproof2}.

So, let us consider a real scalar field $\varphi$ with a mass that varies in time, propagating in a globally hyperbolic and ultrastatic spacetime $\mathbb{I} \times \Sigma$, where $\mathbb{I}$ is any time interval and $\Sigma$ is a Riemannian compact space of three or less dimensions with metric $h_{ij}$ (spatial indexes are denoted with Latin letters from the middle of the alphabet). The field obeys an equation of the KG type (\ref{KG}), where $\Delta$ is the LB operator on $\Sigma$. The mass function $m(t)$ is allowed to be quite arbitrary, assuming only that it has a second derivative which is integrable in any compact subinterval of $\mathbb{I}$. We can construct the canonical phase space of the system $\Gamma$ from the set of Cauchy data given at any reference time $t_0\in \mathbb{I}$: $(\varphi, P_{\varphi}) = (\varphi,\sqrt{h}\varphi^{\prime})|_{t_0}$, equipped with the symplectic structure $\Omega$ that is determined by the standard Poisson brackets $\{\varphi(t_0,\vec{x}), P_{\varphi}(t_0,\vec{y})\} =\delta(\vec{x} - \vec{y})$, for all $\vec{x},\vec{y} \in \Sigma$. Among the infinite number of complex structures \cite{wald} available to determine the Fock representation, we select the complex structure $J_0$ that is naturally associated with the free massless scalar field case, and which can be defined exclusively from the spatial metric, so that it results to be invariant under its symmetries. The properties of the LB operator allow us to decompose the field in a series expansion $\varphi=\Sigma_{nl}q_{nl}\Psi_{nl}$, in terms of a complete set of real eigenmodes $\{\Psi_{nl}\}$ (which provide a basis of the space of square integrable functions on $\Sigma$, with the volume element defined by $h_{ij}$). The discrete eigenvalues of these modes will be called $\{-\omega^2_{n}\}$, with $n\in \mathbb{N}$. It is clear that the degrees of freedom of the field reside in the set of real mode coefficients $\{q_{nl}\}$, which vary only in time. The spectrum of the LB operator may be degenerate. The label $l$ takes this degeneracy into account and runs from 1 to $g_n$, the dimension of the corresponding eigenspace. This degeneracy is always a finite number for compact topologies. The modes satisfy decoupled differential equations of motion:
\begin{equation}
q^{\prime\prime}_{nl}+[\omega^2_n + m^2(t)]q_{nl}=0;
\end{equation}
hence, the time evolution is the same for all modes in the same eigenspace (indicated by the label $n$).

In order to discuss whether the dynamics admits a unitary implementation with respect to the Fock representation determined by $J_0$, it is convenient to define (for all nonzero modes \cite{zero}) the annihilation variables
\begin{equation}
\label{aniquilacion}
a_{nl} = \frac{1}{\sqrt{2\omega_n}}(\omega_n q_{nl} + ip_{nl}),
\end{equation}
together with the corresponding set of creationlike variables given by their complex conjugates $a^*_{nl}$. They provide a new (over-)complete set of variables for the phase space. The action of the complex structure $J_0$ on them is diagonal, with the standard form $J_0(a_{nl})=ia_{nl}$ and $J_0(a^*_{nl})=-ia^*_{nl}$. In other words, $a_{nl}$ and $a^*_{nl}$ can be regarded as the variables that are promoted to annihilation and creation operators in the Fock representation determined by $J_0$. In terms of these pairs of complex variables, time evolution from an initial time $t_0$ to another time $t\in\mathbb{I}$ is given by a Bogoliubov transformation of the form:
\begin{equation}
\label{Evolution}
a_{nl}(t) = \alpha_n(t,t_0) a_{nl}(t_0) + \beta_n(t,t_0) a^*_{nl}(t_0),
\end{equation}
which is block diagonal, owing  to the decoupling of the modes, and insensitive to the degeneracy label $l$ (the dynamics is the same for all modes in the same eigenspace, characterized by $n$). Since this transformation is canonical, the Bogoliubov coefficients satisfy
\begin{equation}
\label{bogoliubov}
|\alpha_n(t,t_0)|^{2} - |\beta_n(t,t_0)|^{2} = 1, \qquad \forall n
\end{equation}
and for every $t \in \mathbb{I}$, independently of the value of $t_0$.

Generally, a canonical transformation $U$ is implementable quantum mechanically as a unitary operator in the Fock representation determined by a complex structure $J$ if and only if the antilinear part of the transformation, given by $U_J=\frac{1}{2} (U+JUJ)$, is a Hilbert-Schmidt operator \cite{SH}. This condition implies that the antilinear coefficients of the considered transformation must be square summable. Namely, for the evolution transformation, the condition is $\sum_n g_n|\beta_n(t,t_0)|^2<\infty$, taking into account the degeneracy of each of the eigenspaces. Clearly, the summability depends on the asymptotic behavior of $\beta_n$ (and $g_n$) in the ultraviolet. The analysis of the asymptotic properties of the spectrum of the LB operator (when $\omega_n \to \infty$) shows that the leading term in the beta function is proportional to $\omega_n^{-2}$. It then turns out that the unitarity condition is equivalent to the finiteness of $\sum_n g_n \omega_n^{-4}$. On the other hand, the number of eigenstates whose eigenvalue does not exceed $\omega^2$ in norm is known to grow in $d$ dimensions at most like $\omega^d$. Hence, one can prove that the above condition is satisfied, in fact, for all Riemannian compact manifolds in three or less dimensions. In other words, the chosen free massless Fock representation supports a unitary implementation of the dynamics defined by Eq. (\ref{KG}), even if the  field has a time dependent mass.

Once proven the existence of a Fock quantization compatible with both the spatial symmetries and a unitary dynamics, the next question is whether this quantization is unique or there exist other physically distinct ones with the same properties. To characterize the most general complex structure that is invariant under the group of spatial symmetries, it is necessary to appeal to a suitable application of Schur's lemma \cite{schur}. The action of the group of symmetries on the canonical phase space is naturally unitary and commutes with the LB operator. Hence, one can decompose the phase space into a convenient sum of finite dimensional subspaces, so that the invariant complex structures adopt (in the basis of massless annihilation and creationlike variables) a $2\times2$ block diagonal form with identical blocks in each subspace (and generically different blocks for different subspaces). As a result, one can easily check that every invariant complex structure $J$ can be related with $J_0$ by a symplectic transformation $K$ in the way $J=KJ_0K^{-1}$, where $K$ admits the same decomposition in blocks that has been found for $J$. We can express each of these $2\times2$ matrix blocks in terms of two complex numbers that can be seen as Bogoliubov coefficients and satisfy the corresponding symplectomorphism condition (\ref{bogoliubov}). At this point, the unitary equivalence between the two complex structures $J$ and $J_0$ can actually be proven by assuming the unitary implementation of the time evolution in the representation determined by $J$. The proof employs that the evolution is unitarily implementable in the whole of $\mathbb{I}$ (and involves a subtle time average over a finite time interval \cite{generalproof1}). Therefore, we conclude that all invariant complex structures which, in addition, allow for a unitary dynamics are indeed equivalent, showing the validity of our criterion to pick out (essentially) a unique Fock representation.

Another issue that has been investigated in detail concerns the possibility of choosing an alternate field description, related with the original one by a time dependent canonical transformation, a possibility which arises naturally in the context of field theories in nonstationary spacetimes. The most general linear canonical transformation that is considered includes a scaling of the field configuration and permits a contribution proportional to the field in the new momentum. Following the line of reasoning of Ref. \cite{generalproof2}, one can show that the criterion of spatial invariance and unitary time evolution also removes this ambiguity and determines a unique canonical pair for the field among all those related by the mentioned transformations. The main points in the proof of this result are based on the calculation of the Bogoliubov coefficients associated with the dynamics of the new annihilation and creationlike variables, introduced in terms of the original ones, and of the coefficients corresponding to the canonical transformation ($K$) between two invariant complex structures. In this manner, one can demonstrate that no complex structure exists admitting a unitary implementation of the dynamics for any of the considered canonical pairs, unless the scaling function for the field configuration is the unit constant function. There still remains the freedom of a redefinition of the momentum by means of a linear contribution of the field. However, and depending on the dimensionality of the space, it turns out that this contribution must either be zero (for two or three spatial dimensions) or (in one spatial dimension) it leads to a quantization which is unitarily equivalent to our fiducial quantization. In this sense, the uniqueness of the field description is ensured.

\section{Generalization of the field equations}

In this section, we will explore how to extend the application of the above uniqueness results to many more physical systems. As we have commented, we are interested in providing a reliable procedure for the quantization of classical fieldlike systems, removing the intrinsic ambiguities of QFT in nonstationary backgrounds. The existence of ambiguities in the construction of a quantum field description affects all branches of Physics. However, the problem is especially relevant in the case of cosmology because the windows available for observable quantum phenomena are indeed narrow, if any, and our Universe is the only accessible system. Therefore, providing criteria that select privileged quantum theories and ensure the robustness of the corresponding physical predictions seems crucial. It is only in this way that one can reach significant quantum predictions, which one may try to compare with the available observational data in order to confirm or falsify the model.

With this motivation in mind, let us start by considering a real scalar field $\phi$ which is subject to the dynamical equation:
\begin{equation}\label{general}
\phi'' + c(t)\phi' - d(t) \Delta\phi + \widetilde{m}^2(t)\phi =0.
\end{equation}

This is the most general second-order differential equation of KG type with coefficients depending (only) on time, provided that the spatial variation enters the dynamics only via the LB operator. In order to relate it to the field equation (\ref{KG}), we allow for two different kinds of transformations. On one hand, we can always introduce a time reparametrization, $T(t)$, given by a bijection on the considered time interval $\mathbb{I}$. In the previous section, when we investigated the unitary implementation of the dynamics, we only analyzed finite transformations between two instants of time, but we did never consider infinitesimal transformations. So, it is fairly straightforward to see that, if the evolution from an initial time $t_0$ to any time $t$ is implemented by a unitary operator $\hat{U}(t, t_0)$, there exists a unitary operator which implements the evolution in the reparametrized time, from $T_0=T(t_0)$ to $T=T(t)$, namely:
\begin{equation}
\hat{U}(T, T_0) = \hat{U}\left(t(T), t(T_0)\right).
\end{equation}
On other hand, as we have already said, in nonstationary contexts one often scales the field configuration by time varying functions.

In total, hence, we introduce a scaling of the field and a time reparametrization given by
\begin{equation}\label{changes}
\phi(t,\vec{x}) = f(t)\varphi(t,\vec{x}), \qquad \mathrm{d}T=g(t) \mathrm{d}t.
\end{equation}
The reparametrization is well defined if $g(t)$ is positive ---or negative--- definite in $\mathbb{I}$ (and integrable in each compact subinterval).
We now substitute these relations in Eq. (\ref{general}), and divide it by the coefficient of the second-time derivative, to set this coefficient equal to the unity. Then, if we force the term that goes with the first-time derivative of the field to disappear and the function that multiplies the LB operator to be constant, as we find in Eq. (\ref{KG}), we obtain [assuming that the function $g(t)\neq0$, according to our previous comments]
\begin{eqnarray}\label{reparametrizacion}
g(t)&=& s \sqrt{d(t)}, \qquad s=\pm,\\
\label{rescalado}
f(t) &=& C d(t)^{-1/4} \exp{\left[-\frac{1}{2} \int^t c(\tilde{t}) \mathrm{d}\tilde{t}\right]},
\end{eqnarray}
where $C\neq 0$ is a constant that we include for convenience, rather than absorb it into the lower integration limit for $c(t)$. Consequently, we see that it is always possible to change to a field formulation with the desired properties. Indeed, this transformation is unique, apart from a possible time reversal (i.e., the selection of the sign $s$). Summarizing, we can find a unique transformation ---obtained by combining a scaling of the field and a time reparametrization--- that leads to a new description which satisfies the field equation (\ref{KG}), regardless of the mass function of the original field $\phi$. And, as we have already proven, this new formulation admits a Fock representation, which is essentially unique, that respects the invariance under the spatial symmetries and the unitary implementation of the dynamics.

The new mass function of the field $\varphi$ is given by
\begin{equation} \label{masa}
m^2(t) =\frac{\widetilde{m}^2(t)}{d(t)} -\frac{d^{\prime\prime}(t)}{4d^2(t)} + \frac{5[d^{\prime}(t)]^2}{16d^3(t)} - \frac{c^{\prime}(t)}{2d(t)} - \frac{c^2(t)}{4d(t)}.
\end{equation}
According to this expression, the new mass explodes [$m^2(t) \to \infty$] if the function $d(t)$ vanishes at any instant $t$. Furthermore, we also see in Eq. (\ref{reparametrizacion}) that other complications arise when $d(t)$ becomes negative, as the time $T$ turns imaginary. In the light of these subtleties, we can identify three kinds of situations, depending on the sign of $d(t)$. When the function $d(t)$ is positive, generically, our scaling of the field and change of time is unique and well defined. When the function $d(t)$ is negative, our analysis requires an extension in order to allow for imaginary times, but the formulas that we have deduced continue to be valid once this extension is done. Finally, when $d(t)$ vanishes, the relations that provide the scaling of the field and the time reparametrization become ill defined, and the mass \eqref{masa} generally becomes singular if $\widetilde{m}$ is finite. These issues will be discussed in more detail later on. Let us also comment that, with our change in time reparametrization and scaling, there is no guarantee a priori that the transformed squared mass $m^2$ be positive. However, this poses no obstruction for the application of our uniqueness criterion \cite{generalproof1,generalproof2}.

On other hand, as we mentioned in the previous section, the new mass must have a second derivative that is integrable in every compact subinterval of the time domain. This is the case if the same condition is true for the mass of the original field, and the functions $c(t)$ and $d(t)$ have a third and a fourth derivative, respectively, which are integrable in all compact time subintervals, once we have assumed that the function $d$ has a definite sign.

As we have explained in Sec. II, it is fundamental to perform the appropriate transformations on the canonical pair of field variables in order to reach the privileged one that admits a Fock quantization with spatial symmetry invariance and a unitary dynamics. To conclude the present section, we will analyze the part of the time dependent canonical transformation that affects the canonical field momentum, to complete the expression of this privileged pair. We start from the most general expression for the original field momentum allowed by linearity. This amounts to requiring linearity on the configuration variable and its time derivative, with time dependent coefficients:
\begin{equation}
P_\phi = \sqrt{h}( A\phi + B\dot{\phi}).
\end{equation}
Notice that the momentum has been densitized. Here, the dot denotes the derivative with respect to the new time $T$, and $A$ and $B$ are two arbitrary functions of time. In the Introduction, we also commented an example of how essential the selection of the canonical momentum is, and how a wrong choice can lead to erroneous conclusions \cite{edd,dadeSit}. In order to apply our uniqueness results, we must ensure that the conjugate momentum of the new field $\varphi$ is $P_\varphi = \sqrt{h} \dot{\varphi}$. Using this requirement and demanding the transformation to be canonical, we deduce the possible form of $B$: $B = f^{-2}$. Finally, this provides the general expression allowed for $P_{\phi}$ and the transformed momentum:
\begin{eqnarray}
P_\phi &=& \sqrt{h}\left( A\phi + \frac{1}{f^2}\dot{\phi}\right), \\
P_\varphi &=& f P_\phi - \sqrt{h}\left( f A + \frac{\dot{f}}{f^2}\right)\phi.
\end{eqnarray}

\section{Spacetime interpretation of the field equations}

In order to supply an interpretation of our field equations as generalized KG equations of a scalar field propagating in a nonstationary background, let us consider conformally ultrastatic spacetimes (i.e. with a constant-norm timelike Killing vector) with normal spatial sections, whose metric can be expressed as
\begin{equation}\label{confspacet}
ds^2 = -N^2(t)\mathrm{d}t^2 + a^2(t) h_{ij}(x)\mathrm{d}x^i \mathrm{d}x^j,
\end{equation}
where $N^2(t)$ is the lapse function arising from a $3+1$ decomposition of the metric \cite{ADM}, $a(t)$ is the scale factor, and $h_{ij}$ are the spatial components of the metric.

Calculating the d'Alembertian associated with this metric and substituting it into the KG equation for a scalar field with a time dependent mass $\bar{m}(t)$, we obtain
\begin{equation}\label{ecum}
\phi''+\left[\ln\left(\frac{a^3(t)}{N(t)}\right)\right]' \phi' -\frac{N^2(t)}{a^2(t)}\Delta\phi +[N(t)\bar{m}(t)]^2\phi =0.
\end{equation}

We can compare this expression with Eq. (\ref{general}) and determine the variable functions of the metric (\ref{confspacet}) in terms of the functions $c(t)$ and $d(t)$, obtaining
\begin{eqnarray}\label{factorescala}
a^4(t) &=& d(t) \exp\left[\int^t 2c(\tilde{t})\mathrm{d}\tilde{t}\right], \\
\label{lapso}
N^4(t) &=& d^3(t) \exp\left[\int^t 2c(\tilde{t})\mathrm{d}\tilde{t}\right].
\end{eqnarray}
Therefore, the entire family of second-order equations (\ref{general}) is in one-to-one correspondence to the family of KG field equations in curved spacetimes of the form (\ref{confspacet}).

Moreover, the identification between both families of equations allows us to relate the two mass terms, leading to
\begin{equation}
\tilde{m}^2(t) = [N(t)\bar{m}(t)]^2.
\end{equation}
Namely, the mass associated with Eq. (\ref{general}) is proportional to the lapse function and the mass of the KG field.

It is evident from these expressions that when the function $d(t)$ is zero, the scale factor $a(t)$, the lapse $N(t)$, and hence the mass function $\tilde{m}(t)$ vanish as well. As an additional consequence, the lapse tends to zero faster than the scale factor, according to the dependence of both quantities on $d(t)$ displayed in Eqs. (\ref{factorescala}) and (\ref{lapso}). Another relevant result that follows from this comparison is that, in such scenarios, when the time reparametrization (\ref{reparametrizacion}) is introduced, the desired scaling (\ref{rescalado}) is the inverse of the scale factor $a(t)$ (apart from a multiplicative constant $C\neq 0$), as one might have expected: $f(t)=C/a(t)$.

At this stage, we can rewrite the line element (\ref{confspacet}) in terms of the functions $c(t)$ and $d(t)$:
\begin{eqnarray}\label{metrica}
ds^2 &=& [-d(t)\mathrm{d}t^2 + h_{ij}(x)\mathrm{d}x^i \mathrm{d}x^j] a^2(t) \\ \nonumber
	 &=& [-d(t)\mathrm{d}t^2 + h_{ij}(x)\mathrm{d}x^i \mathrm{d}x^j] \sqrt{d(t)}\exp\left[\int^t c(\tilde{t})\mathrm{d}\tilde{t}\right],
\end{eqnarray}
in order to analyze the extent to which our arguments can be applied and the validity of the spacetime interpretation of the field equations.

In case the function $d(t)$ has a zero at a given point $t$, we find that the metric degenerates completely and the spacetime interpretation becomes meaningless: in principle, we cannot say anything about what happens at that moment. Furthermore, when one computes the scalar curvature $R$ of the considered metric, one sees that it explodes as $d^{-7/2}$. Motivated by the recent interest on the possible transition from a hyperbolic regime to an elliptic one in LQC \cite{cambiosignatura,deformedLQC,MS,cai}, it is particularly important to analyze the behavior of the Ashtekar-Barbero variables \cite{GCL} when $d(t)$ vanishes. Let us recall that these variables are an $su(2)$ connection and a densitized triad, both canonically conjugate. Using them, one can construct an alternative description of general relativity, with a formulation similar to that of gauge theories \cite{GCL}. In the classical theory, the Ashtekar-Barbero connection is the sum of the extrinsic curvature \cite{waldRG} in its triadic form and of the spin $su(2)$ connection compatible with the triad \cite{LQCmena}. Then, if one calculates the expressions of these variables for the metric (\ref{metrica}), one easily realizes that the densitized triad scales as $a^2(t)$ and, therefore, it also degenerates completely when $d(t)$ becomes zero. Much more important is the behavior of the connection. Although the spin connection of the spatial triad does not depend on the function $d(t)$, the extrinsic curvature (in triadic form) diverges as $d^{-9/4}(t)$. Thus, the zero of $d(t)$ defines a genuine singularity of the theory, where the fundamental variables of LQG are not well defined. This analysis clarifies the discussion in the literature and corrects some misunderstandings about what type of signature change occurs when the sign of the function that multiplies the LB operator flips in Eq. (\ref{general}).

On other hand, if $d(t) < 0$, the metric becomes Euclidean. In order to avoid complex metrics, we can rewrite the scale factor as
\begin{equation}
a^2(t) = \tilde{C} \sqrt{|d(t)|}\exp\left[\int^t_{t_d} c(t)\mathrm{d}t\right].
\end{equation}
Here, $\tilde{C}$ is a positive constant that appears when one fixes the lower limit in the integral, $t_d$, as the point where the function $d(t)$ vanishes, once the square root of the sign of the function $d(t)$ has been absorbed.

As we commented in the Introduction, this process can be regarded as a signature change, from a region of Lorentzian spacetime (with signature $ - + + +$) to a Euclidean region ($+ + + +$). In Eq. (\ref{general}), we already saw that the transition to a Euclidean sector also involves a change in the behavior of the field equation, from a hyperbolic to an elliptic regime. In this transition, the singularity $d(t)=0$ separates the spacetime into two regions with very different nature. In the next section, we will show how, under certain continuity requirements on the solutions of the field equations, it is possible to establish a relationship between the evolution in both regions.

\section{Vacuum dynamics with a signature change}

We want to study how a vacuum state specified by initial conditions in a Euclidean spacetime region evolves to a Lorentzian region when the sign of the function $d(t)$ changes. A physical motivation comes again from the current status in LQC, where it has been argued that the dynamical equations for cosmological perturbations might be affected by a change of this type \cite{cambiosignatura,deformedLQC,MS,cai,ucai}. However, the analysis has important applications in other fields of Physics; for instance, signature changes have already been studied in the context of Bose-Einstein condensates \cite{thesis}. In such circumstances, there may exist reasonable requirements to select a vacuum state in the Euclidean region (where the field equation is elliptic). It then seems meaningful and useful to discuss how these requirements can be translated into conditions for the (evolved) vacuum state in the Lorentzian region, in which the uniqueness of the quantization is guaranteed by our criterion of spatial invariance and unitarity.
Our strategy will be based on imposing certain continuity conditions at the event of signature change, when the scale factor has been shown to vanish and the spacetime metric degenerates. First, we will discuss the vacuum dynamics and the continuity conditions. We will see that some cosmological particle production is generated in the process of signature change. We finally estimate this particle production in the WKB approximation \cite{wkb}.

\subsection{Matching conditions}

Since the evolution equation of the field $\varphi$ is not well defined when the function $d(t)$ vanishes [the mass term in Eq. (\ref{masa}) explodes], we will adopt the description in terms of the KG field $\phi$. We can choose the lapse function as $N^2 = \varepsilon a^6$, where $\varepsilon$ is a parameter that coincides with the sign of $d(t)$, so that $\varepsilon = +1$ corresponds to the Lorentzian region whereas the Euclidean region corresponds to $\varepsilon = -1$. Then, Eq. (\ref{ecum}) becomes
\begin{equation}\label{KGsignatura}
\ddot{\phi} = -\varepsilon \left[a^4 \Delta \phi +a^6 m^2 \phi\right].
\end{equation}
Here, the dot denotes the derivative with respect to the time $\tau$ selected by the lapse function. From expressions (\ref{changes}) and (\ref{factorescala}), it is not difficult to see that
\begin{equation}
\mathrm{d}T^2 = \varepsilon a^4 \mathrm{d}\tau^2 = d(t) \mathrm{d}t^2.
\end{equation}
Furthermore, we assume that the time origin is fixed to coincide with the moment of signature change.

Next, we pick out a complete set of solutions $\{\varphi_n^{\pm}(T)\psi_n(\vec{x})\}$ to the equation (\ref{KG}), where the superscripts $+$ and $-$ indicate the positive and negative frequency solutions, respectively, and $\psi_n$ are eigenfunctions of the LB operator with eigenvalue $-\omega_n^2$. In this section, we ignore the possible degeneracy of the eigenspaces of the LB operator, obviating the corresponding degeneracy label. Besides, we assume that the considered solutions are orthonormalized with respect to the KG product \cite{wald}. These solutions can be understood as those defined from any of the complex structures within our unique unitary equivalence class of invariant structures, which has been selected by the criterion of spatial invariance and unitarity in the evolution of the field $\varphi$ in the Lorentzian regime. We reparametrize this set in terms of the time $\tau$ and scale the functions which form it by the inverse of the scale factor, so as to obtain a basis of solutions for the KG field $\phi$ in the $\varepsilon =1$ sector (i.e., in the Lorentzian regime). We call $\{\phi_n^{\pm} (\tau)\psi_n (\vec{x})\}$ this new set of solutions. In order to attain solutions in the Euclidean regime, we carry out a Wick rotation of these modes, assuming that the analytic continuation of the solutions to imaginary times is well defined. In this way, we obtain a set of solutions for the $\varepsilon = -1$ region, given by
\begin{equation}\label{soleuclidea}
\phi_{n}^{\pm (E)}(\tau)=\lim_{\tilde{\tau}\rightarrow i \tau} \phi^{\pm}_{n}(\tilde{\tau}).
\end{equation}

Any solution to the field equations can be expanded in terms of this set of modes, both in the Lorentzian region (with $\varepsilon = 1$) and in the Euclidean region (with $\varepsilon=-1$) reached by the analytic continuation introduced above (in both cases, with solutions extended to the largest interval allowed by the field equation for $\phi$). We call these expansion coefficients  $c_n^{\pm}$ and $c_n^{\pm(E)}$, respectively.

A vacuum state can be characterized as a specific complex solution of the field equations which includes only the positive frequencies selected by a certain choice of complex structure. This solution may be specified, for instance, giving conditions on a spacetime section. Suppose that, in our case, the conditions fix the coefficients $c_n^{\pm(E)}$ in the region of Euclidean behavior. We can adopt conditions at a time $\tau_0 < 0$ such that $c_n^{+(E)} = 1$ and $c_n^{-(E)} = 0$. These conditions determine the evolution in the Euclidean region up to $\tau = 0$. At that point, it is necessary to provide matching conditions relating the considered solution for the field to another one in the Lorentzian region beyond the signature change. In order to do this, we adopt the ideas presented by Dray et al. \cite{Dray}, which require continuity for the KG field and its time derivative on the hypersurface where the signature flips. An extended analysis can also be found in Ref. \cite{thesis}. We note that this continuity is natural according to equation (\ref{KGsignatura}), which shows no singular behavior in the analyzed process of signature change .

The proposed matching conditions amount to a linear system of equations between the expansion coefficients of the Euclidean modes and those of the Lorentzian modes. For each mode, we get
\begin{eqnarray}\label{sistema}
&&\left(\begin{array}{cc} \phi^{+(E)}_{n}(0) & \phi^{-(E)}_{n}(0)  \\
{\dot{\phi}^{+(E)}}_{n}(0)  & {\dot{\phi}^{-(E)}}_{n}(0) \end{array}\right)\left(
\begin{array}{c} c_n^{+(E)} \\
c_n^{-(E)}\end{array}\right) \nonumber \\
&&=\left(
\begin{array}{cc} \phi^{+}_{n}(0) & \phi^{-}_{n}(0)  \\
{\dot{\phi}^{+}}_{n}(0)  & {\dot{\phi}^{-}}_{n}(0) \end{array}\right)\left(
\begin{array}{c} c_n^{+} \\
c_n^{-}\end{array}\right),
\end{eqnarray}
where the evaluation at zero denotes the limit in which the time approaches the moment of signature change.

In the Lorentzian region, the solutions are {\sl orthonormalized} with respect to the KG-like product \cite{note2}:
\begin{equation}
\langle \phi, \chi \rangle= \frac{\dot{\phi}\chi - \phi\dot{\chi}}{i}.
\end{equation}
In fact, it can be seen that this property of orthonormalization is valid for both formulations: the field $\phi$ in time $\tau$ and $\varphi$ in time $T$. Specifically, $\langle \phi^{-}_n, \phi^{+}_m \rangle = \delta_{nm}$, while the products of two solutions of positive or negative frequencies cancel. Using this and defining
\begin{equation} \label{matrixelements}
I_n^{(rs)}=\lim_{\tau\rightarrow 0} \langle \phi^{r (E)}_{n}(-|\tau|), \phi^{(s)}_{m}(|\tau|) \rangle,
\end{equation}
we obtain the solution to the system (\ref{sistema}) as
\begin{eqnarray}\label{solsistema}
\left(
\begin{array}{c} c_n^{+} \\
c_n^{-} \end{array} \right) =
\left(\begin{array}{cc} -I_n^{(+-)} & -I_n^{(--)}  \\
I_n^{(++)}  & I_n^{(-+)} \end{array} \right)\left(
\begin{array}{c} c_n^{+ (E)} \\
c_n^{- (E)} \end{array}\right).
\end{eqnarray}

If we specialize the discussion to the choice of vacuum determined by $c_n^{+(E)} = 1$ and $c_n^{-(E)} = 0$, that is, by the absence of negative frequencies initially, we see that the solution resulting for the vacuum in the Lorentzian region has coefficients $c_n^{+} = -I_n^{(+ -)}$ and $c_n^{-} =I_n^{(++)}$. Since these coefficients correspond to contributions of positive and negative frequencies in the Lorentzian regime, respectively, they can be interpreted as Bogoliubov coefficients in the sense of providing the components that either preserve the positive frequencies or instead mix them with the negative ones. Then, this latter mixing term predicts the production of particles with respect to the initial state ($c_n^{-} \neq 0$).

Finally, for completeness, we construct the field $\varphi$ which presents a unitary evolution in the region of Lorentzian behavior:
\begin{equation}
\varphi = a(T) \sum_n \left\{\left(c_n^{+} \phi^{+}_n [t(T)] + c_n^{-} \phi^{-}_n [t(T)]\right) \psi_n(\vec{x})\right\}.
\end{equation}

\subsection{WKB approximation: Particle production}

In the ultraviolet sector ($\omega_n \gg 1$), the mass term can be neglected in Eq. (\ref{KGsignatura}) in comparison to the LB term (for finite values of the scale factor), and we may apply the WKB method \cite{wkb} to get an estimate of the particle production associated with the process of signature change.

We start with the approximate solutions of positive and negative frequency in the Lorentzian regime:
\begin{equation}\label{approsol}
\phi_n^{\pm} \simeq  \frac{\exp\left[\pm i\omega_n\int_{\tau_0}^{\tau} a^2(\tilde{\tau})  \mathrm{d}\tilde{\tau}\right]}{\sqrt{2\omega_n a^2(\tau)}},
\end{equation}
where the exponent has been computed integrating from an initial time $\tau_0$ to any time $\tau$. For the Lorenztian region, we can set this initial time equal to zero. On the other hand, using Eq. (\ref{soleuclidea}) we can determine the solution in the Euclidean region by analytic continuation (see also Ref. \cite{wkb}). If we make correspond $\tau_0$ in the Euclidean region with the point at which we fix the initial vacuum state, we have $|\tau_0| > |\tau|$ in that sector. Notice that $|\tau|$ decreases as we approach the signature change hypersurface (located where $\tau$ vanishes).

Employing the WKB solutions for the field and its first time-derivative on both sides of that hypersurface and taking \emph{appropriately} the limit $\tau \to 0$, we can calculate the elements $I_n^{(rs)}$ of the matrix that appears in Eq. (\ref{solsistema}), given by expression (\ref{matrixelements}). In this way, we obtain
\begin{equation}
I_n^{(rs)}= -\frac{e^{r\omega_n \Lambda}}{2} (s+ir), \qquad \Lambda= \int^{|\tau_0|}_0 \bar{a}^2(\tilde{\tau}) \mathrm{d}\tilde{\tau}.
\end{equation}
Here $\bar{a}^2$ is the analytic continuation of the squared scale factor under a Wick rotation and a time reversal [$\bar{a}^2(-\tau)=\lim_{\tilde{\tau}\rightarrow i \tau}a^2(\tilde{\tau})$].

Summarizing, if we provide initial conditions for the vacuum state in the Euclidean region so that it only contains contributions from positive frequencies, the resulting Lorentzian coefficients $c_n^{-}$ indicate an exponentially amplified particle production, owing to the vacuum dynamics in the process of signature change. The amount of particle production depends on the characteristics of the Euclidean region only through the quantity $\Lambda$.

\section{Conclusions}

We have investigated the validity and the range of applicability of an approach recently introduced in QFT in curved spacetimes to remove the ambiguities in the Fock quantization of scalar fields in nonstationary backgrounds. In this work, our main motivation has been the potential applications to cosmological spacetimes. The uniqueness criterion put forward in this approach consists in demanding invariance under the group of spatial symmetries of the field equations and requiring the unitary implementability of the dynamics. First, we have considered the case of a KG field with a time varying mass propagating in (ultra-)static backgrounds with compact spatial sections of three or less dimensions. The relevance of this case is already warranted by the fact that it can be regarded as a rescaled KG field in a nonstationary (cosmological) spacetime. We have seen that the Fock representation that is naturally associated with the free massless case allows an invariant quantization with unitary evolution even for massive fields. After characterizing all invariant complex structures, one can show that those which allow a unitary evolution are (unitarily) equivalent to the free massless one, so that the infinite ambiguity in the choice of a Fock representation is eliminated in the quantization. In addition we have seen that, to attain the desired quantum spatial invariance and unitary dynamics, there only exists one possible scaling of the field and (except in the case of one spatial dimension) a unique choice of its canonical momentum within all the canonical pairs related by linear time dependent canonical transformations. Thus, the proposed criterion also removes the ambiguity in the selection of a canonical pair among all those connected by these transformations. A detailed demonstration of these results can be found in Refs. \cite{generalproof1,generalproof2}. The particular case of compact flat spatial topology, which deserves a careful study owing to some specific peculiarities and given its importance for physical applications (recall the apparent flatness of the observed universe), has been analyzed in Ref. \cite{jcaptoro}.

We have then shown that the unitarity of the evolution does not depend on the time reparametrization used in the description of the system (as far as the change of time parameter is a bijection). Taking into account this property and allowing the scaling of the field by time dependent functions, we have extended the range of applicability of our criterion to all evolution equations of second order in time derivatives of the generalized KG type (\ref{general}), in which the spatial dependence appears only through a LB term. Any equation of this kind can be reformulated as that of a KG field on a static background, but with a time dependent mass, by introducing a scaling of the field and a suitable choice of time. As a matter of fact, both the field scaling and the choice of time parameter are determined in a unique way (up to the possibility of a time reversal). The only necessary condition (apart from certain requirements of differentiability on the time dependent coefficients of the field equation) is that the coefficient of the LB term is a function with well defined sign in the studied time domain (a positive function if a hyperbolic evolution equation is expected).

In our analysis, we have also shown that the considered family of field equations (to which we have extended the uniqueness criterion) admit all the interpretation of a KG equation for a scalar field with variable mass in a conformally ultrastatic spacetime. The choice of the lapse and scale factor of this spacetime determines, univocally, the coefficients of the first time derivative of the field and of the LB term in the differential equation. If this latter coefficient becomes negative, the field equation becomes elliptic and can be understood as corresponding to a Euclidean spacetime region. Therefore, the change of sign in the coefficient of the LB term can be viewed as indicating a transition from a Lorentzian to a Euclidean region, in a process of signature change. However, it is important to emphasize that, as we have demonstrated, when this change occurs the spacetime metric degenerates completely. So, the discussed phenomenon of signature change involves a singularity in our system, if one adheres to the spacetime interpretation of the field equations, singularity where the curvature invariants diverge. Moreover, given the applications of our results in the context of LQC, we have computed the corresponding Ashtekar-Barbero variables, proving that they are ill defined at the singularity, namely, on the hypersurface associated with the signature change. Consequently, strictly speaking, any spacetime interpretation is lost there and the classical description in terms of the fundamental variables of LQC becomes meaningless on that hypersurface.

We have also commented that our discussion is partly motivated by the present status of deformed covariance algebras in LQC \cite{holonomy,inverse,both,cai,ucai}, which suggests the emergence of a Lorentzian universe like ours from a Euclidean spacetime at high-energy regimes \cite{deformedLQC}. This has opened a debate about the possibility of setting the initial conditions for cosmological states in a Euclidean spacetime region ---or instead just on the hypersurface corresponding to the signature change \cite{asilence}---. We have proposed a formalism to deal with initial conditions formulated in Euclidean regimes, showing how they can be translated into initial conditions in the Lorentzian sector by means of the vacuum evolution and suitable continuity conditions in the process of signature change, making the framework compatible with the selection of a Fock quantization based on the uniqueness criterion that we have put forward. The continuity conditions that we have suggested are inspired by the works of Dray et al. \cite{Dray}, in which particle production is proven to occur for a massless scalar field propagating in a signature changing spacetime, consisting of a Euclidean region between two Lorentzian parts of the spacetime. In addition, an extensive discussion about signature changing events and their consequences in Bose-Einstein condensates can be found in Ref. \cite{thesis}. Following these ideas, in the last part of our work we have dealt with the issue of a possible cosmological particle production in the process of signature change. We have demonstrated that, if the initial conditions select a vacuum with no contribution of negative frequencies (after analytic continuation) in the Euclidean region, particles are indeed produced. Moreover, using the WKB method to approximate the solutions in the ultraviolet sector, we have found an exponential production in terms of the integral of the square scale factor over the Euclidean region.

Finally, let us emphasize that this particle production is due to the quantum evolution of the vacuum in the Euclidean regime, a phenomenon which is apparently independent of the amplification of modes \cite{amplifications} that has also been confirmed recently in the effective description of inhomogeneous models in hybrid LQC \cite{fmo}, systems in which the cosmological background always remains Lorentzian and a bounce in the expansion occurs.

\section*{Acknowledgements}

We would like to thank C. Barcel\'o, L. Garay, and E. Wilson-Ewing for discussions and comments. This work was supported by the Project No.\ MICINN/MINECO FIS2011-30145-C03-02 from Spain.


\begin{thebibliography}{99}
\bibitem{note1} In the absence of the density corrections arising from LQC,  this equation dictates the evolution in conformal time. We keep the discussion more general in the sense that the time $t$ may be, in principle, any time parameter. Later on we will discuss the relation of this parameter with a time coordinate in a conformally ultrastatic spacetime if the Mukhanov-Sasaki equation is to be regarded as a KG equation with a time dependent mass.
\bibitem{zero} The consideration of a finite number of modes, like e. g. the zero modes, does not change the conclusions about the unitary implementability of the dynamics in the Fock quantization of the field.
\bibitem{note2} More precisely, the solutions are orthonormal with respect to the product $\langle \phi, \chi \rangle_{KG}=\langle \phi^{*}, \chi \rangle$.
\end{thebibliography}
\end{document}